\documentclass[journal, twocolumns]{IEEEtran}

\usepackage[utf8]{inputenc}
\usepackage{epsfig, amsthm, amsmath, amssymb, amsfonts, subfigure,color}

\usepackage{cite}
\usepackage{graphicx}
\usepackage{stfloats}
\usepackage[table,xcdraw]{xcolor}
\usepackage{tikz}
\usepackage{enumitem}
\usepackage{url}
\usepackage{array}
\usepackage{multirow}
\usepackage{booktabs} % for better looking tables
\usepackage{siunitx}  % for units of measure and data in tables

\usepackage{xfrac}
\usepackage{nicefrac}
\usepackage{psfrag}
\usepackage{times}
\usepackage{balance} 
\usepackage{amssymb}
\usepackage{amsfonts}
\usepackage{bm}
\usepackage[bottom]{footmisc}
\usepackage{mwe}
\usepackage{color}
\usepackage{verbatim}
\usepackage{graphicx}
\usepackage{booktabs,multirow}
\usetikzlibrary{patterns,arrows}
\usepackage{epstopdf}
\usepackage{hyperref}

\usepackage{algorithm}
\usepackage{algorithmicx}
\usepackage{algpseudocode}
\usepackage{tabularx}

\usepackage[acronym]{glossaries}

\newacronym{iid}{i.i.d.}{independent and identically distributed}

\newacronym{lqr}{LQR}{linear quadratic regulator}

\newacronym{bs}{BS}{base station}

\newacronym{resnet}{ResNet}{residual network}

\newacronym{awgn}{AWGN}{additive white Gaussian noise}

\newacronym{gpr}{GPR}{Gaussian process regression}

\newacronym{inf}{InF-SH}{Indoor Factory}

\newacronym{std}{std}{standard deviation}

\newacronym{ofdma}{OFDMA}{orthogonal frequency division multiple access}

\newacronym{prb}{PRB}{physical resource block}

\newacronym{jepa}{JEPA}{joint-embedding predictive architecture}

\newacronym{i-jepa}{I-JEPA}{Image-Joint-Embedding Predictive Architecture}

\newacronym{v-jepa}{V-JEPA}{Video-Joint-Embedding Predictive Architecture}

\newacronym{ts-jepa}{TS-JEPA}{time-series joint-embedding predictive architecture}

\newacronym{h-jepa}{H-JEPA}{Hierarchical Joint-Embedding Predictive Architecture}

\newacronym{mcs}{MCS}{modulation and coding schemes}

\newacronym{snr}{SNR}{signal-to-noise ratio}

\newacronym{cdf}{CDF}{cumulative distribution function}

\newacronym{pdf}{PDF}{probability distribution function}

\newacronym{ctde}{CTDE}{centralized training with decentralized execution}

\newacronym{mse}{MSE}{mean squared error}

\newacronym{los}{LoS}{line-of-sight}

\newacronym{nlos}{NLoS}{non-line-of-sight}

\newacronym{pca}{PCA}{principal component analysis} 

\newacronym{sgd}{SGD}{stochastic gradient descent} 

\newacronym{nrmse}{NRMSE}{normalized root mean squared error} 

\newacronym{cnns}{CNNs}{convolution neural networks} 

\newacronym{ebms}{EbMs}{energy-based models} 

\newacronym{jea}{JEA}{joint-embedding architecture} 

\newacronym{ga}{GA}{Generative Architecture} 

\newacronym{vit}{ViT}{Vision Transformer} 

\newacronym{ema}{EMA}{exponential moving average} 

\newacronym{mlp}{MLP}{Multi-layer Perceptron} 

\newacronym{cnn}{CNN}{Convolutional Neural Network } 

\newacronym{relu}{ReLu}{rectifier linear unit} 

\newacronym{mae}{MAE}{mean absolute error}

\newacronym{mape}{MAPE}{mean absolute percentage error}

\newacronym{nmae}{NMAE}{ normalized mean absolute error}

\newacronym{t-sne}{t-SNE}{t-Distributed Stochastic Neighbor Embedding}

\newacronym{urllc}{URLLC}{ultra-reliable low-latency communication}

\newacronym{mmtc}{mMTC}{massive machine-type communication}

\newacronym{ai}{AI}{artificial intelligence}

\newacronym{ml}{ML}{machine learning}

\newacronym{rnns}{RNNs}{recurrent neural networks}

\newacronym{lstm}{LSTM}{long short-term memory}

\newacronym{gru}{GRU}{gated recurrent unit}

\newacronym{aos}{AoS}{age of semantics}

\newacronym{aoi}{AoI}{age of information}

\newacronym{byol}{BYOL}{bootstrap your own latent}

\newacronym{sc}{SC}{Semantic Communication}

\newacronym{vae}{VAE}{variational auto-encoder}

\newacronym{simclr}{SimCLR}{simple framework for contrastive learning of visual representations}

\newacronym{bpp}{bpp}{bits per pixel}

\newacronym{arima}{ARIMA}{autoregressive integrated
moving average}

\newacronym{csi}{CSI}{channel state information}

\newacronym{hmpc}{HMPC}{Hierarchical Model Predictive Control}

\definecolor{color1}{RGB}{237, 191, 193}
\definecolor{color2}{RGB}{229, 153, 157}
\definecolor{color3}{RGB}{225, 123, 116}
\definecolor{color4}{RGB}{217,87,77}
\definecolor{color5}{RGB}{203,52,38}

\def\l{\left}
\def\r{\right}
\def\({\l(}
\def\){\r)}
\def\[{\l[}
\def\]{\r]}

\def\BibTeX{{\rm B\kern-.05em{\sc i\kern-.025em b}\kern-.08em
    T\kern-.1667em\lower.7ex\hbox{E}\kern-.125emX}}

\begin{document}

\title{Hierarchical JEPA Meets Predictive Remote Control in Beyond 5G Networks}

\author{
Abanoub M. Girgis,~\IEEEmembership{Student Member,~IEEE},
Ibtissam~Labriji,~\IEEEmembership{Member,~IEEE},
and Mehdi~Bennis,~\IEEEmembership{Fellow,~IEEE}

%\thanks{This work was supported in part by the RCF-Korea  (Semantics-Native Communication and Protocol Learning in 6G); and in part by the Research Council of Finland (former Academy of Finland) Project Vision-Guided Wireless Communication.}

\thanks{A. M. Girgis and M. Bennis are with the Center for Wireless Communications, University of Oulu, Oulu 90014, Finland (e-mail: abanoub.pipaoy@gmail.com; mehdi.bennis@oulu.fi).}
\thanks{I. Labriji is with Nokia Bell Labs, Massy, France (e-mail: ibtissam.labriji@nokia-bell-labs.com).}
}

\maketitle

\begin{abstract}
In wireless networked control systems, ensuring timely and reliable state updates from distributed devices to remote controllers is essential for robust control performance. 
However, when multiple devices transmit high-dimensional states (e.g., images or video frames) over bandwidth-limited wireless networks, a critical trade-off emerges between communication efficiency and control performance.
To address this challenge, we propose a \textit{Hierarchical Joint-Embedding Predictive Architecture} (H-JEPA) for scalable predictive control. 
Instead of transmitting states, device observations are encoded into low-dimensional embeddings that preserve essential dynamics. 
The proposed architecture employs a three-level
hierarchical prediction, high-level, medium-level, and low-level predictors operating across different temporal resolutions, to achieve long-term prediction stability, intermediate interpolation, and fine-grained refinement, respectively.   
Control actions are derived within the embedding space, removing the need for state reconstruction. 
Simulation results on inverted cart-pole systems demonstrate that H-JEPA enables up to $42.83 \%$ more devices to be supported under limited wireless capacity without compromising control performance.  
\end{abstract}

\begin{IEEEkeywords}
self-supervised learning, hierarchical, joint-embedding predictive architecture, semantic communication.
\end{IEEEkeywords}

\section{Introduction}
\label{Sec_Introd}
\IEEEPARstart{D}{evelopment} of sixth-generation networks entails a paradigm shift towards intelligent wireless networks capable of supporting emerging applications such as autonomous vehicles~\cite{parekh2022review} and smart factories~\cite{chen2017smart}. 
These applications will strain the capacity of the current networks, primarily due to the high-dimensional nature of transmitted data, such as images and video frames. 
This comes in tandem with the rapid advancement in artificial intelligence (AI) and its remarkable impact on communication and control systems, driving the need for AI-native and control-aware network design. 
Hence, the convergence of control and AI is becoming essential for future wireless networks.  
The issue of optimizing wireless networks for remote control of distributed devices has gained significant attention in the literature.  
Typically, sensors transmit their raw states to remote controllers, which then compute and transmit control actions to satisfy control tasks.  
Recent approaches employ deep learning to improve communication efficiency through extracting and transmitting low-dimensional representations instead of raw device states~\cite{girgis2024semantic}. 
While these approaches reduce redundant information and improve task-specific efficiency, the learned representations lack semantic meaning and generalization capability across different tasks. 
Moreover, they suffer from error accumulation over long-term prediction horizons, as repeated use of latent dynamics for multi-step prediction degrades control accuracy and scalability. 

Another prominent approach introduces predictive control models that capture device dynamics in the raw state space~\cite{girgis2021predictive,yamak2019comparison,hua2019deep}. 
These models predict future states, enabling proactive decision-making to reduce control latency or improve reliability under limited network capacity. 
However, their prediction accuracy degrades significantly over long horizons, limiting their applicability to short-term tasks such as next-frame video prediction~\cite{elbamby2018toward}.  
This is because their dependence on single-scale temporal dynamics leads to error accumulation and instability for long-term predictions.

To address these limitations, we introduce a self-supervised \textit{\gls{hmpc} framework} that enables the controller to learn device dynamics from high-dimensional visual observations. 
This \gls{hmpc} framework comprises high-level, medium-level, and low-level predictors operating at different temporal resolutions to ensure long-term prediction stability, intermediate interpolation, and fine-grained refinement, respectively. 
Instead of modeling raw sensory inputs, the proposed framework learns device dynamics within a low-dimensional joint-embedding space, capturing the essential latent dynamics in a hierarchical architecture. 
Within the embedding space, a semantic actor model is utilized to map embeddings to control actions, eliminating the need to reconstruct high-dimensional visual observations.   

The most comparable works~\cite{girgis2024time,chaaya2025pixels} focus on vision-based remote control and utilize joint-embedding predictive architectures for low-level latent state prediction to maintain control performance under limited network capacity. 
Unlike Dreamer~\cite{hafner2019dream}, which trains a generative world model through reinforcement learning and optimizes an expected-return objective, the proposed framework adopts a discriminative joint-embedding objective that directly minimizes a control-aligned loss in the latent space, ensuring tighter coupling between prediction quality and control performance.   
Our proposed framework extends beyond these approaches by enhancing prediction accuracy across multiple hierarchical temporal levels, while reducing communication overhead and maintaining control performance. 
Simulation results demonstrate that the proposed framework supports $42.83\%$ more devices under $20 \, \mathrm{dB}$ signal-to-noise ratio (SNR) compared to baselines. 

\section{System Model}
\label{Sec_System}

\begin{figure}[t]
    \centering
    \includegraphics[width=0.48\textwidth]{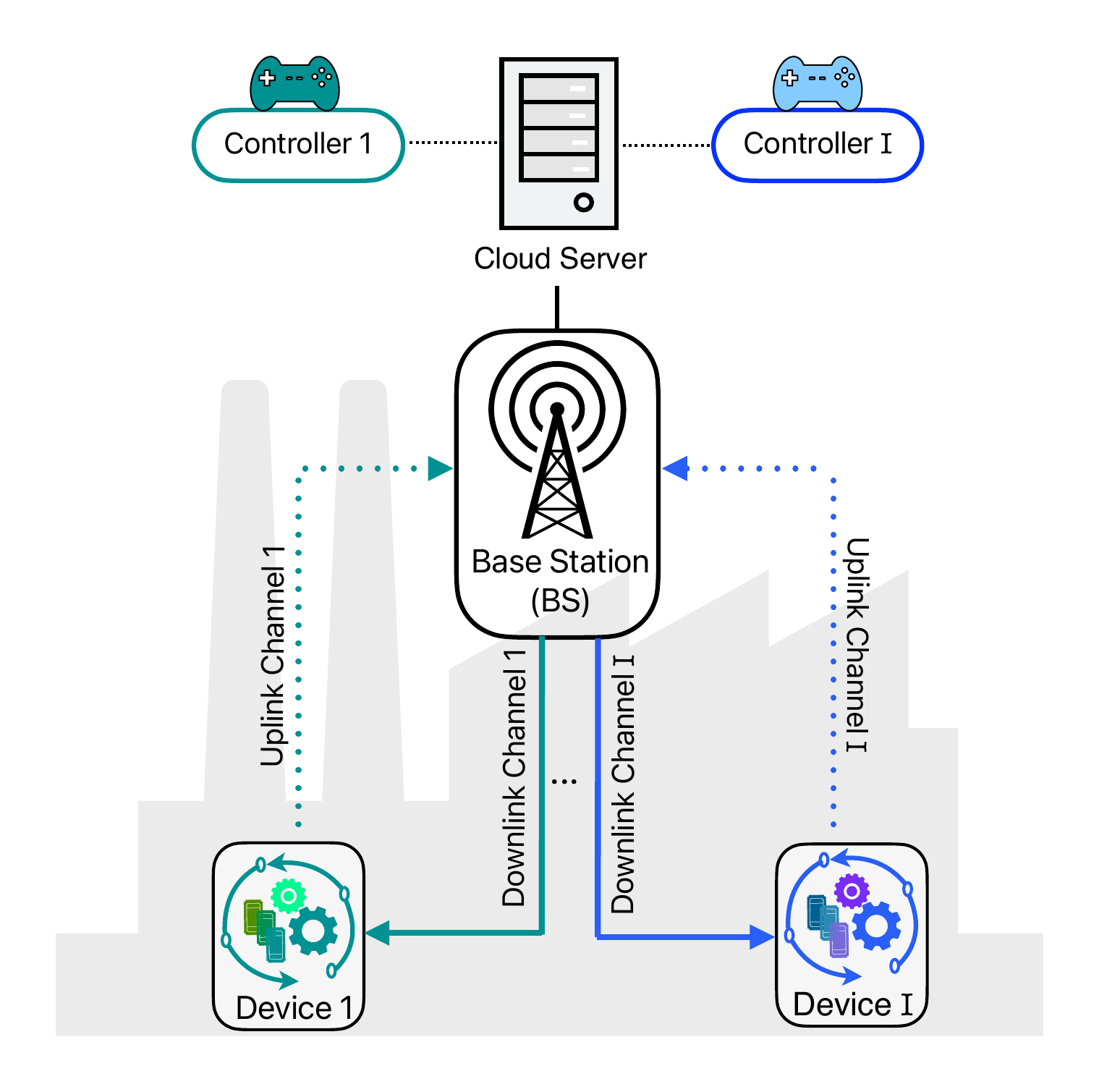} \\
    \caption{A wireless networked control system in a smart factory.}
    \label{fig_system_model}
\end{figure}

We consider a wireless networked control system, as illustrated in Fig.~\ref{fig_system_model}, where multiple independent non-linear control loops share the wireless channels. 
Each loop consists of a device, a co-located sensor-actuator pair, and a remote controller. 
The device embeds a non-linear dynamic \textit{process} whose state needs to be regulated towards a desired operating point. 
The \textit{actuator} applies control actions, while the \textit{sensor} periodically samples the device state and transmits it to the remote controller via \textit{uplink transmission}. 
Upon receiving the device state, the \textit{controller} computes the corresponding control action and transmits it back to the actuator via downlink transmission, closing the control loop.

\subsection{Control System}
\label{Sec_System_A}
Each device in the wireless networked control system is indexed by $i \in \mathcal{I}$, with $\mid \mathcal{I} \mid$ denoting the total number of devices. 
At time $t= k \tau_{o}$, the sensor of device $i$ samples a $p$-dimensional state vector $\mathbf{x}_{i,k} \in \mathbb{R}^{p}$ at a fixed sampling rate $\tau_{o}$, which is transmitted to the remote controller. 
Based on $\mathbf{x}_{i,k}$, the controller computes a target $q$-dimensional control action $\mathbf{u}_{i,k} \in \mathbb{R}^{q}$ that is transmitted back to the actuator over ideal channels. 
The device state evolution follows a discrete-time non-linear dynamics given as~\cite{ogata2010modern}
\begin{align}
    \label{eq1_non_linear_evol}
    \mathbf{x}_{i,k+1} = \mathbf{f}_{i} \left( \mathbf{x}_{i,k} , \mathbf{u}_{i,k} \right) + \mathbf{n}_{s,k},
\end{align} 
where $\mathbf{n}_{s,k} \sim \mathcal{N} \left( \mathbf{0}, N_{s} \mathbf{I} \right)$ represents \gls{iid} Gaussian process noise. 
The function $\mathbf{f}_{i}: \mathbb{R}^{p} \times \mathbb{R}^{q} \rightarrow \mathbb{R}^{p}$ captures non-linear dynamics of device $i$. 

The remote controller aims to compute the optimal actions using the non-linear control policy in~\cite{khalil2002nonlinear}.  
As illustrated in Fig.~\ref{fig_system_model}, each control loop is closed through wireless channels to ensure successful control performance. 
Hence, reliable uplink transmission is critical since packet losses prevent the appropriate actuation and cause $\mathbf{x}_{i,k} \rightarrow \infty$ as $k \rightarrow \infty$. 

\subsection{Wireless Communication System}
\label{Sec_System_B}

We consider a sparse clutter and a high base station Indoor Factory (InF-SH) scenario, where the devices are randomly distributed within the coverage area of \gls{bs}~\cite{etsi5138,9312675}. 
The controllers are assumed to be located on a distant cloud server with negligible communication delay between the cloud server and~\gls{bs}. 
The devices and \gls{bs} communicate through wireless channels. 
The channel between the device and \gls{bs} follows a standard path loss and Rayleigh block fading~\cite{goldsmith2005wireless}, where the channel gain remains constant over the time interval of $\tau_{o}$ but varies independently across different time intervals. %
The received \gls{snr} at time $t = k \tau_{o}$ is given as~\vspace{-8pt} 
\begin{align}
    \label{eq_SNR}
    \gamma_{i,k} = 10^{-\frac{\mathrm{PL}^{\mathrm{NLoS}}_{\mathrm{dB}}}{10}} \frac{P \mid H_{i,k} \mid^{2} }{ N_{c}}, 
\end{align}
where $P$ denotes the fixed transmission power, $H_{i,k}$ represents the Rayleigh flat-fading channel between the device $i$ and \gls{bs} at time $k$, $\mathrm{PL}^{\mathrm{NLoS}}_{dB}$ represents the non-line-of-sight path loss, and $N_{c}$ denotes the additive white Gaussian noise power. 
The uplink channel capacity for device $i$ at time $k$ is given as~\vspace{-4pt}
\begin{align}
    \label{eq_UL_Rate}
    R_{i,k} = W_{i} \log_{2} (1 + \gamma_{i,k}),
\end{align}
where $W_{i}$ denotes the allocated transmission bandwidth.  

A transmission outage occurs when $R_{i,k}$ falls below a predefined threshold $\bar{R}$. 
Then, the transmission outage probability is given as \vspace{-6pt}
\begin{align}
    \label{eq_outage}
    \epsilon_{i,k} &= \mathbb{P} \left[ R_{i,k} < \bar{R} \right] \nonumber \\ &=  1 - \text{exp} \left[ -10^{\frac{\mathrm{PL}^{\mathrm{NLoS}}_{\mathrm{dB}}}{10}} \frac{ N_c }{P} \left(2^{ \frac{\bar{R}}{W_{i} } } -1\right)  \right], 
\end{align}
which depends on channel conditions, transmission power, and allocated bandwidth. 
Hence, efficient utilization of wireless resources is fundamental to minimizing transmission failures and ensuring reliable uplink transmission for maintaining robust control performance.  

\subsection{Problem Statement}
\label{Sec_System_C}

In wireless networked control systems, maintaining timely and reliable state updates (camera frames) from multiple devices to remote controllers remains a challenge to scalable control under limited wireless resources.    
To address this challenge, we propose a novel \textit{\gls{hmpc} framework} to ensure efficient long-horizon prediction. 
Specifically, instead of transmitting raw states, the device state $\mathbf{x}_{i,k}$  is first mapped into low-dimensional embeddings $\mathbf{z}_{i,k} = \Psi(\mathbf{x}_{i,k})$, which capture the essential details to obtain latent dynamics. 
Building on these embeddings, we introduce a hierarchy of three predictive mappings designed to operate across different temporal resolutions. 
At the highest level, a \textit{high-level predictor} $\mathcal{P}^{H}(\cdot)$ predicts long-horizon embeddings. 
To bridge these broader intervals, a \textit{medium-level predictor} $\mathcal{P}^{M}(\cdot)$ interpolates intermediate-scale embeddings between successive high-level predictions.
Finally, at the finest granularity, the \textit{low-level predictor} $\mathcal{P}^{L}(\cdot)$ refines the embeddings further by interpolating low-level embeddings within medium-level embeddings. 
This hierarchical multi-timescale prediction mitigates error accumulation over long-horizon prediction, improving scalable and robust control under limited wireless resources. 

\begin{figure}[t]
    \centering
   \includegraphics[trim=0 35 0 35, clip, width=1.02\linewidth]{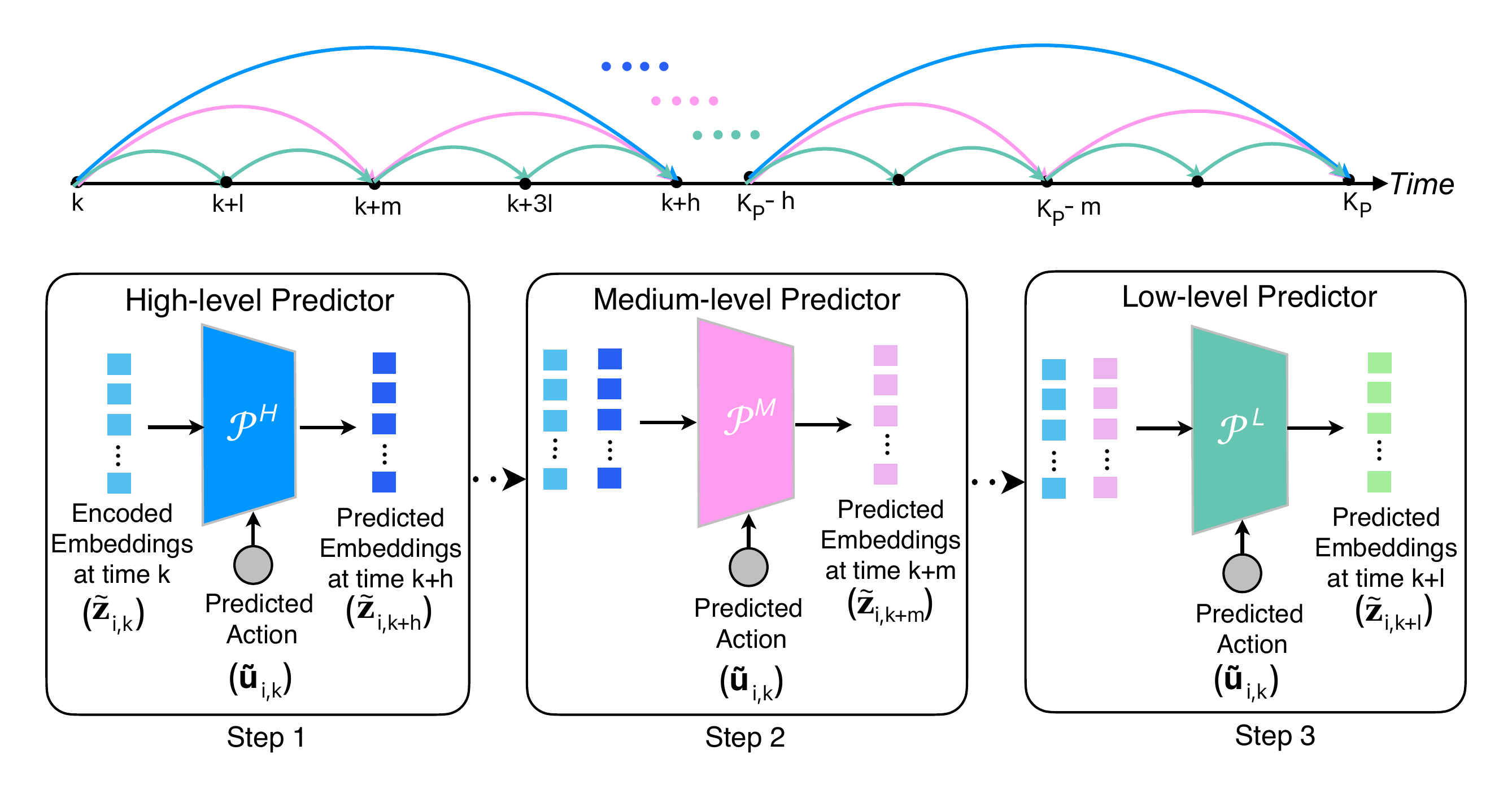} \\
   \caption{The hierarchical model predictive control.}
    \label{fig_predictor}
\end{figure}

\section{Hierarchical Model Predictive Control}
\label{Sec_solution}

To address the challenges of communication efficiency and scalability in the wireless networked control system, we propose a novel hierarchical model predictive control framework that integrates a \textit{\gls{h-jepa}} with a \textit{semantic actor model}.

\subsection{Hierarchical Joint-Embedding Predictive Architecture}
\label{Ch5_1}
The proposed \gls{h-jepa} is a self-supervised hierarchical predictive technique that efficiently encodes high-dimensional device states into low-dimensional semantic embeddings while decomposing the long-term prediction horizon into three hierarchical levels to capture long-term and fine-grained latent dynamics. 
The key components of \gls{h-jepa}, as illustrated in Fig.~\ref{fig_predictor}, are described below.
\begin{enumerate}
    \item \textit{Context Encoder:} At time step $t= k \tau_{o}$, the context encoder $\Psi_{\theta}(\cdot)$, parameterized by a set of $\theta$ learnable parameters, maps the high-dimensional device state $\mathbf{x}_{i,k}$ into low-dimensional embeddings given as
    $\mathbf{z}_{i,k} = \Psi_{\theta}(\mathbf{x}_{i,k})$.

    \item \textit{Target Encoder:} For a defined prediction horizon $K_{p}$, the target encoder $\Psi_{\bar{\theta}} (\cdot)$ processes a sequence of future high-level device states $( \mathbf{x}_{i,k+h}, \dots, \mathbf{x}_{i,k+K_{p}} )$ into corresponding low-dimensional semantic embeddings $( \mathbf{z}_{i,k+h}, \dots, \mathbf{z}_{i,k+K_{p}} )$, serving as prediction targets. 
    The parameters $\bar{\theta}$ are updated via an \gls{ema} of the context encoder to ensure stable training. 
    The target encoder's parameters update rule is given as~\vspace{-4pt} 
    \begin{align}
        \label{eq_weights_update1}
         \bar{\theta} \leftarrow \eta \bar{\theta}  + (1 - \eta) \theta. 
    \end{align} 
    Specifically, the target encoder mirrors the context encoder’s architecture and shares identical parameters at initialization. 
    Since the target encoder provides labels for training the context encoder and predictors, its gradients are blocked through its branch to prevent representation collapse, and its weights are updated using an \gls{ema} of the context encoder's parameters in~\eqref{eq_weights_update1}. 
    %
    %This iterative process gradually enhances the context encoder's ability to produce meaningful representations from which the high-level predictor infers high-level future representations. 
    %

    \item \textit{High-level Predictions:} The high-level predictor $\mathcal{P}^{H}_{\varphi_{H}}$, parameterized by a set of  $\varphi_{H}$ learnable parameters, captures the non-linear high-level embedding evolution to predict target high-level embeddings from current embeddings and predicted actions$( \tilde{\mathbf{u}}_{i,k}, \dots, \tilde{\mathbf{u}}_{i,k+K_{p}-h} )$, solving an auto-regressive task given as~\vspace{-6pt}
    \begin{align}
       \tilde{\mathbf{z}}_{i,k+sh} = \mathcal{P}^{H}_{\varphi_{H}} \left( \mathbf{z}_{i,k}| \tilde{\mathbf{u}}_{i,k}, \dots, \tilde{\mathbf{u}}_{i,k+K_{p}-h} \right), \label{eq_pred5}
    \end{align} 
    with $s \in [1, \frac{K_{p}}{h} ]$, where $h$ is the high-level target prediction depth, $\tilde{\mathbf{z}}_{i,k+h}$ is the high-level predicted embeddings at time $ t= (k+h) \tau_{o}$, and $\tilde{\mathbf{u}}_{i,k}$ is the predicted action of device $i$ at time $t = k \tau_{o}$.
    The parameters of the high-level predictor $\varphi_{H}$ and the context encoder $\theta$ are jointly learned using a cosine similarity loss to align the high-level predicted embeddings with their target embeddings through gradient-based optimization as~\vspace{-4pt} 
    \begin{align}
         \label{eq_opt_high}
         \underset{ \theta, \varphi_{H} }{\arg \min}  \; \; \frac{1}{K_{p}} \sum_{k = 1}^{K_{p}} \frac{ \langle  \tilde{\mathbf{z}}_{i,k+h}, \mathbf{z}_{i,k+h} \rangle}{ \|  \tilde{\mathbf{z}}_{i,k+h} \|_{2}.  \|  \mathbf{z}_{i,k+h}) \|_{2}},     
    \end{align}
    while the target encoder's parameters $\bar{\theta}$ are updated, at each training step, via an \gls{ema} of the context encoder's parameters.

    \item \textit{Medium-level Predictions:} The medium-level predictor $\mathcal{P}^{M}_{\varphi_{M}}$, parameterized by a set $\varphi_{M}$ of learnable parameters, captures the non-linear evolution of medium-level embeddings to interpolate embeddings between the current and high-level predicted embeddings, conditioned on current predicted control actions $( \tilde{\mathbf{u}}_{i,k}, \dots, \tilde{\mathbf{u}}_{i,k+K_{p}-m} )$, solving an auto-regressive task as~\vspace{-4pt} 
    \begin{align}
       \tilde{\mathbf{z}}_{i,k+s'm} 
         = \mathcal{P}^{M}_{\varphi_{M}} \left( \mathbf{z}_{i,k}^{m}\,\big|\, \tilde{\mathbf{u}}_{i,k}, \dots, \tilde{\mathbf{u}}_{i,k+K_{p}-m} \right)
          \label{eq_pred6}
    \end{align}
    with $s' \in [1, \frac{K_{p}}{m}  ]$, where  $m < h$ is the medium-level target prediction depth, $ \mathbf{z}_{i,k}^{m} = [\mathbf{z}_{i,k},\tilde{\mathbf{z}}_{i,k+h}]$ is the concatenated current and high-level embeddings, and $\tilde{\mathbf{z}}_{i,k+m}$ is the medium-level predicted embeddings at time $ t= (k+m) \tau_{o}$.
    The medium-level predictor parameters $\varphi_{M}$ are learned using a cosine similarity loss, defined analogously to~\eqref{eq_opt_high}, to align the medium-level predicted embeddings with their corresponding targets.

    \item \textit{Low-level Predictions:} The low-level predictor $\mathcal{P}^{L}_{\varphi_{L}}$, parameterized by a set  $\varphi_{L}$ of learnable parameters, captures the non-linear evolution of low-level embeddings to interpolate embeddings between current and medium-level predicted embeddings, conditioned on current predicted control actions $( \tilde{\mathbf{u}}_{i,k}, \dots, \tilde{\mathbf{u}}_{i,k+K_{p}-l} )$, solving an auto-regressive task as~\vspace{-4pt}
    \begin{align}
        \tilde{\mathbf{z}}_{i,k+s''l} = \mathcal{P}^{L}_{\varphi_{L}} \left( \mathbf{z}_{i,k}^{l}| \tilde{\mathbf{u}}_{i,k}, \dots, \tilde{\mathbf{u}}_{i,k+K_{p}-l}  \right), \label{eq_pred7}
    \end{align}
    with $s'' \in [1, \frac{K_{p}}{l}  ]$, where $l < m $ is the low-level target prediction depth, $\mathbf{z}_{i,k}^{l} = [ \mathbf{z}_{i,k},\tilde{\mathbf{z}}_{i,k+m}]$ is the concatenated current and high-level embeddings, and $\tilde{\mathbf{z}}_{i,k+l}$ is the low-level predicted embeddings at time $ t= (k+l) \tau_{o} $. 
    The low-level predictor parameters $\varphi_{L}$ are learned using a cosine similarity loss, defined analogously to~\eqref{eq_opt_high}, to align the low-level predicted embeddings with their corresponding targets.
\end{enumerate}

In a nutshell, the proposed \gls{h-jepa} for predictive remote control operates through two distinct phases. 
\begin{enumerate}
    \item \textit{Training Phase}: During this phase, the devices transmit their high-dimensional states to remote controllers for control action computation. 
    Meanwhile, the context encoder along with high-level, medium-level, and low-level predictors are trained at \gls{bs} on datasets consisting of a pair of consecutive states and their corresponding actions $\mathcal{D}_{H} = \{ \mathbf{x}_{i,k+h}, \mathbf{u}_{i,k+h} \}_{k=1}^{K_{p}}$,  $\mathcal{D}_{M} = \{ \mathbf{x}_{i,k+m}, \mathbf{u}_{i,k+m} \}_{k=1}^{K_{p}}$, and $\mathcal{D}_{L} = \{ \mathbf{x}_{i,k+l}, \mathbf{u}_{i,k+l} \}_{k=1}^{K_{p}}$, respectively.   
    The training process minimizes hierarchical prediction errors until convergence. 

    \item \textit{Inference Phase}: Once training is complete, the devices deploy the learned context encoder to locally encode high-dimensional states into low-dimensional embeddings. 
    At the cloud server, the learned high-level predictor predicts high-level embeddings conditioned on predicted actions, while the medium- and low-level predictors conditioned on predicted actions fine-grained high- and medium-level predicted embeddings, respectively. 
    %
    %This hierarchical prediction enables the prediction of latent dynamics at multiple timescales without frequent state transmissions, ensuring efficient scalability under limited wireless resources.
    %
\end{enumerate}

While low-dimensional embeddings reduce communication overhead, they are not directly actionable for control action computation. 
To bridge this gap, the \textit{semantic actor model} introduced in~\cite{girgis2024time} is utilized to map predicted embeddings into calculated actions. 
This ensures that the predicted embeddings can be directly leveraged for closed-loop control, eliminating the need for high-dimensional state reconstruction.

\begin{figure*}
    \centering
    \subfigure[action prediction accuracy.\label{fig_control_pred}]{\includegraphics[width=0.24\textwidth]{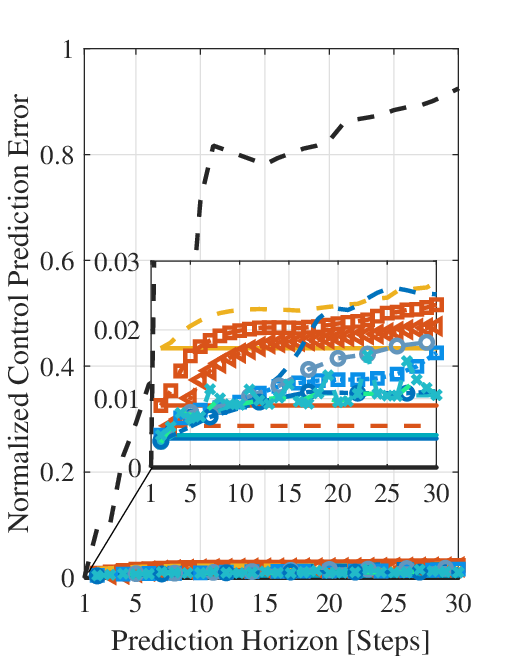}}
    \subfigure[Communication cost.\label{fig_comm_bits}]{\includegraphics[width=0.24\textwidth]{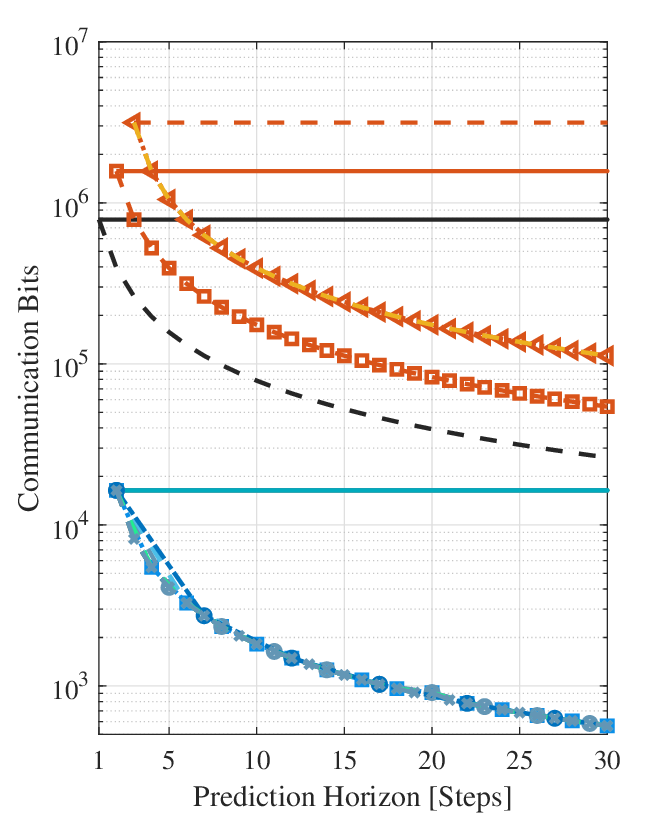}} 
    \subfigure[Control accuracy.\label{fig_norm_score}]{\includegraphics[width=0.42\textwidth]{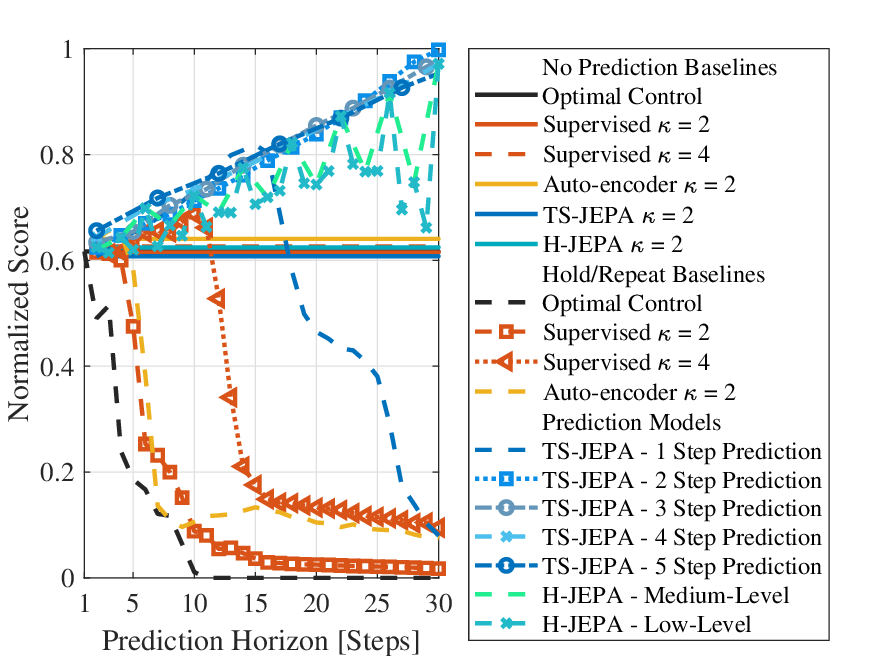}} 
   \caption{Comparison between the proposed hierarchical model predictive control and different baseline models in a single device scenario.}
    \label{fig_H-jepa_pred}
\end{figure*} 

\section{Simulation Results}
\label{Sec_results}

To evaluate the performance of the proposed \gls{hmpc} framework, we conducted simulations on inverted cart-pole systems.
Each system state is represented by an RGB frame sampled at a sampling rate of $\tau_{o} = 1 \, \mathrm{ms}$ over a trajectory length of $100$ time steps, ensuring sufficient temporal resolution. 
The control actions apply horizontal forces to the cart within predefined limits of $u_{max} = 20 \, \mathrm{N}$, and $u_{min} = -20 \, \mathrm{N}$~\cite{morasso2019stabilization}.
The high-level dataset $\mathcal{D}_{H}$ includes $200$ training and $40$ testing trajectories of RGB frames paired with control actions. 
The medium- and low-level datasets $\mathcal{D}_{M}$ and $\mathcal{D}_{L}$ comprise $80$ training and $40$ testing trajectories of low-dimensional embeddings, along with their associated control actions. 
The weight parameters of the proposed framework are trained by minimizing the loss in~\eqref{eq_opt_high} using SGD with a batch size of $256$~\footnote{All simulation hyperparameters and implementation details are available at \url{https://github.com/abanoubpipaoy/Hierarchical-JEPA-Meets-Predictive-Remote-Control-in-Beyond-5G-Networks/tree/simulation_parameters}}. 
The context encoder employs a deep convolutional \gls{resnet} with three layers of $64$, $128$, and $256$ neurons, each followed by batch normalization and \gls{relu} activation function. 
The high-,medium-,and low-level predictors are structured as \gls{mlp} with one hidden layer of $1024$ neurons and $256$-dimensional output. 
For benchmarking, we employ the evaluation metrics and baseline comparisons presented in~\cite{girgis2024time} and follow an
identical structure to~\cite{girgis2024time}.

\noindent \textbf{Encoding and Prediction Capability Evaluation}: Fig.~\ref{fig_H-jepa_pred} compares the proposed \gls{hmpc} framework with the baseline models in a single device scenario, evaluating control action prediction accuracy, communication cost, and control score across the prediction horizon. 
To validate the encoding capability of the proposed framework, we first examine the no-prediction case, where the remote controller receives either high-dimensional states or low-dimensional embeddings at each time step. 
The proposed \gls{h-jepa} with two consecutive frames ($\kappa=2$) is compared against: (i) a supervised model with $\kappa=2$ and $\kappa=4$, (ii) an auto-encoder model with $\kappa=2$, (iii) the \gls{ts-jepa} with one-step prediction, and (iv) the optimal non-linear control policy.  
As shown in Fig.~\ref{fig_control_pred}, the proposed \gls{h-jepa} achieves a normalized control prediction error close to the optimal non-linear policy while requiring significantly fewer communication bits (Fig.~\ref{fig_comm_bits}). 
The encoder of the proposed \gls{h-jepa} performs comparably to the \gls{ts-jepa}, as both share the same encoder design, confirming their ability to extract semantic embeddings from high-dimensional states.   
This results in a near-optimal normalized control score, as shown in Fig.~\ref{fig_norm_score}. 
In contrast, the auto-encoder model suffers from high normalized control prediction error due to reconstruction errors and limited temporal data, while the supervised model struggles with generalization, requiring more consecutive frames $\left(\kappa = 4 \right)$ to improve control performance, albeit with increased communication cost.

To evaluate the prediction capability of the proposed framework, we restrict the device to transmit only at the initial time step, while the remote controller predicts future actions.    
Here, the proposed framework is compared to baseline models employing repeated-action and zero-action strategies, as well as the one-step \gls{ts-jepa} prediction.   
The results show that the proposed \gls{h-jepa} achieves the lowest control prediction error over the prediction horizon, outperforming the \gls{ts-jepa} with one-step target depth, which suffers from rapid error accumulation.  
The \gls{ts-jepa}s with longer target prediction depths $\left( 2 \mathsf{-} 5 \; \text{target depths} \right)$ mitigate error accumulation but lack fine-grained details between prediction steps.   
In contrast, the \gls{h-jepa} mitigates both issues by decomposing the prediction horizon into the \textit{high-level}, which is modeled as the \gls{ts-jepa} with a four-step target prediction depth to predict long-term targets, the \textit{medium-level} interpolates between the initial and high-level steps, and the \textit{low-level} interpolates finer details between medium-level prediction steps.   
This hierarchical decomposition enables stable long-term predictions while retaining temporal detail. 

\begin{figure}[t]
    \centering
   \includegraphics[width=0.48\textwidth]{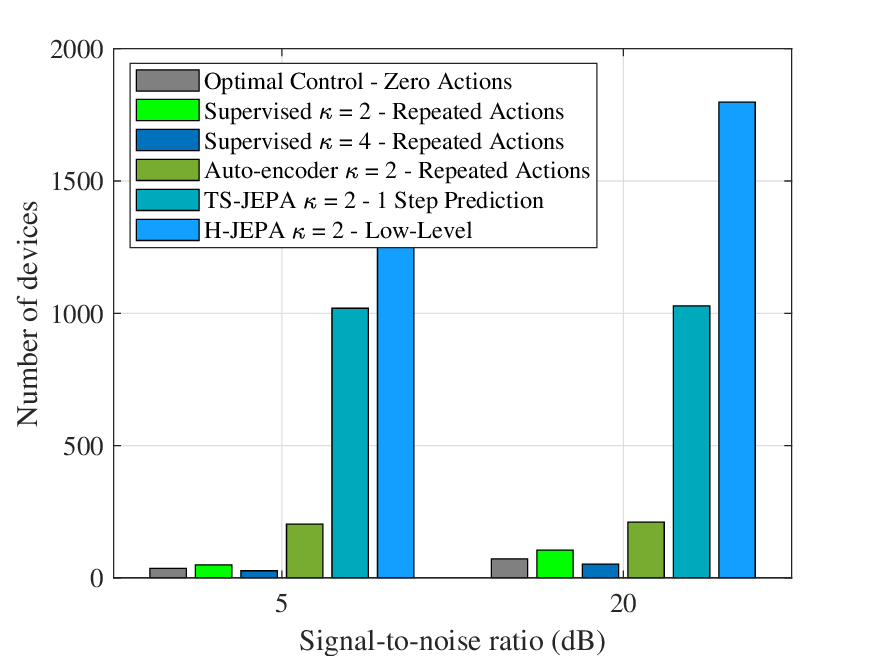} \\
   \caption{Total number of devices under different target SNRs for the proposed hierarchical model predictive control and different baseline models.}
    \label{fig_H_JEPA}
\end{figure} 

\noindent \textbf{System Scalability with Prediction Capability}: Fig.~\ref{fig_H_JEPA} evaluates the scalability of the proposed framework compared to the baseline models using zero-action, repeated-action, and one-step prediction strategies.  
The scalability is measured as the maximum number of devices that can be supported while maintaining acceptable control performance, defined by a normalized control score in the range $\left[ 0.62, 1.0 \right]$ under different \gls{snr}s. 
The results show that the proposed \gls{h-jepa} consistently supports more devices than all baseline models at different \gls{snr}s. 
This improvement arises from two key design features. 
First, the proposed \gls{h-jepa}, like the \gls{ts-jepa}, effectively compresses high-dimensional states into low-dimensional embeddings, reducing transmission overhead and accommodating more devices. 
Unlike the one-step \gls{ts-jepa} prediction, the proposed \gls{h-jepa} decomposes the prediction horizon into hierarchical levels.
This hierarchical strategy not only improves long-term prediction accuracy but also compensates for missing updates when the channel faces adverse channel conditions, maintaining robust control performance without retransmission.  
In contrast, generative models such as supervised learning and auto-encoders exhibit poor scalability due to their reliance on transmitting high-dimensional states, which saturates network capacity, and a lack of predictive capability under adverse channel conditions.   

\section{Conclusion}
\label{Sec_conclusion}

This letter proposes a novel hierarchical self-supervised framework for scalable predictive control under limited wireless resources. 
The proposed framework encodes high-dimensional states into low-dimensional embeddings that capture the essential dynamics with a three-level hierarchical prediction, enabling efficient communication without compromising control performance. 
Simulation results validate the effectiveness of the proposed framework in terms of control scalability under limited wireless resources. 
Building on the promising performance of the hierarchical framework, we will explore its benefits for uncertainty-aware scheduling.  

\bibliographystyle{IEEEtran}
\bibliography{IEEEabrv,bibliography}
\end{document}